# Cycloid motions of aggregates in a dust plasma

FENG Fan, ZHANG Yong-Liang, YAN Jia, LIU Fu-Cheng, DONG Li-Fang, HE Ya-Feng[*]

Hebei Key Laboratory of Optic-electronic Information Materials, College of Physics Science and Technology, Hebei University, Baoding 071002, China.

**Abstract**: Hypocycloid and epicycloid motions of aggregates consisted of one large and one small grains are experimentally observed in a rf dust plasma. The cycloid motions are regarded as combination of a primary circle and a secondary circle. Measurements with high spatiotemporal resolution show that the secondary circle is determined by the initial angle velocity of the dropped aggregate. The primary circle originates from the asymmetry of the aggregate. The small grain in the aggregate always leads the large one as they travelling, which results from the difference of the resonance frequency of the two grains. Comparison experiments with regular microspheres show that the cycloid motions are distinctive features of aggregates immersed in a plasma.

**Keywords**: Dust plasma, aggregate, cycloid motion.

**PACS:** 52.27.Lw

## 1. Introduction

The dust plasmas appear over a range of environments such as the comet tails, the magnetic confinement fusion, and the semiconductor processing. [1–3] The dust particles immersed in a plasma could acquire charges sufficiently large such that they couple strongly and exhibit crystallization and phase transition in the dust subsystem. [4–7] In the laboratory, dust particles are deliberately introduced into a plasma or can form and grow in the plasma due to different physical processes. Spherical dust grains such as the glass, melamine formaldehyde, and alumina mocrospheres are often used in DC or rf discharge experiments. [4, 5, 8] Due to the spherical symmetry, they exhibit perfect structures such as the hexagonal lattice. [4] In theory, the grain is often considered as a charged particle neglecting the grain shape. [9–13] These are very helpful in understanding the processes of condensed matter physics at mesoscale. However, the grains appeared in planetary rings, fusion reactors, materials processing, etc., are typically elongated or aggregates consisting of many small subunits. The dynamics of the nonspherical grain in the dust plasma is less known at present.

Aggregate tends to acquire more charge compared to spherical grains of the same mass due to their porous structure. Cylindrical or elongated dusts also possess nonzero dipole moment. Charge-dipole interaction becomes very import in this case. [14] When the centers of the dipole and the gravity do not coincide the aggregate begins to rotate. For a grain with irregular shape the net transfer of the angular momenta from plasma flux and surrounding gas can be nonzero [15], which gives rise to the spin of the grain. The angular frequency of such spin can reach a rather large values ($\sim 10^4$–$10^9$

s$^{-1}$). [16, 17] The spin of a hollow glass microsphere with defect has been detected with angular velocity up to 12000 rad/s in a stratified glow discharge. [16] Magnus effect originated from the fast spin of an irregular grain results in complex cycloid motions. Here, we report an experimental study on the cycloid motions of aggregates immersed in a rf plasma. Hypocycloid, epicycloid, and elliptical motions of aggregate consisted of one large and one small grains are observed. Based on the measurement with high spatiotemporal resolution the origin of these cycloid motions are analyzed.

## 2. Experimental apparatus

**Figure 1** shows the schematic diagram of the experimental apparatus. The experiments are carried out in a vacuum chamber with a background pressure within p=10-70 Pa. Plasma is produced by coupling capacitively the electrode to a rf generator (13.56 MHz). The forward power varies from 2 to 40 W. The separation between the lower stainless steel electrode and the upper ITO coated electrode is 40 mm. A glass ring is positioned on the lower stainless steel electrode. The diameter and the height of the glass ring are 30 mm and 10 mm, respectively.

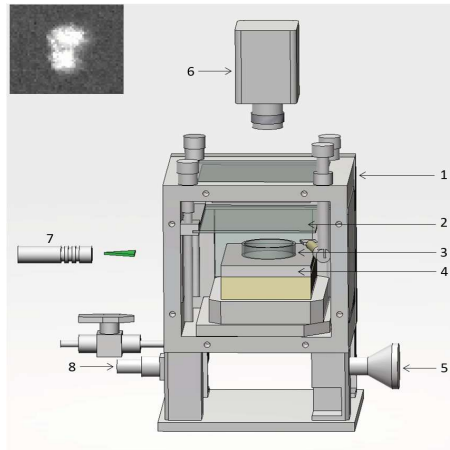

Figure. 1: (Color online) Schematic diagram of the experimental setup. The inset shows the aggregate consisted of one large and one small grains. 1-chamber, 2-upper electrode: ITO, 3-glass ring, 4- lower electrode: stainless steel, 5-vacuum pump, 6-high speed camera, 7-laser sheet, 8-rf power.

Pollen with average radius d=20±5 μm are injected into the plasma through a shaker and serve as dust grains. Few pollens are injected into the plasma during each experiment in order to avoid the formation of grains ring and plasma condensation. Often, two or more pollens could form an aggregate. Here, we will focus on the movement of an aggregate consisted of two pollens (with $d_a$≈25 μm and $d_b$≈15 μm), as shown by the inset in **figure. 1**. After being injected into the plasma, charged aggregate levitates nearby the glass wall. The aggregate is horizontally illuminated by a 50 mW, 532 nm laser sheet with a thickness of 0.5 mm and imaged by a high speed camera (PCO.dimax, 1000 fps, field of view 7×7 mm$^2$) from the upper transparent electrode. The spatial resolution is about 3.8 μm/pixel. Characteristic parameters of the aggregate movement are measured by using the image processing with MATLAB.

## 3. Experimental results

**Figure 2(a)-2(c)** show three types of trajectories of travelling aggregates. These trajectories are superpositions of 300 images captured by the high speed camera. **Figure 2(a)** shows an epicycloid motion with inward petals. This meandering motion can be regarded as a combination of two circular motions, where the primary circle (radius $r_2$) orbits the secondary circle (radius $r_1$) in one direction with angular velocity $\omega_1$ and spins about its center in the same direction with angular velocity $\omega_2$ as shown by the diagram of **figure. 2(d)**. **Figure. 2(c)** shows a hypocycloid motion with outward petals with rotations in the opposite sense [as illustrated by the diagram of **figure. 2(e)**]. For epicycloid trajectory, the direction of the angular velocity of the primary circle is the same as that of the secondary circle, i.e., $\omega_1\omega_2 > 0$. For hypocycloid trajectory, the direction of the angular velocity of the primary circle is opposite to that of the secondary circle, i.e., $\omega_1\omega_2 < 0$. In addition, if the values of the angular velocities are $\omega_1 = 0$, $\omega_2 \neq 0$, the aggregate would follow an elliptical trajectory as show in **figure. 2(b)**. Therefore, the value of $\omega_1\omega_2$ could serve as the order parameter which identifies these trajectories.

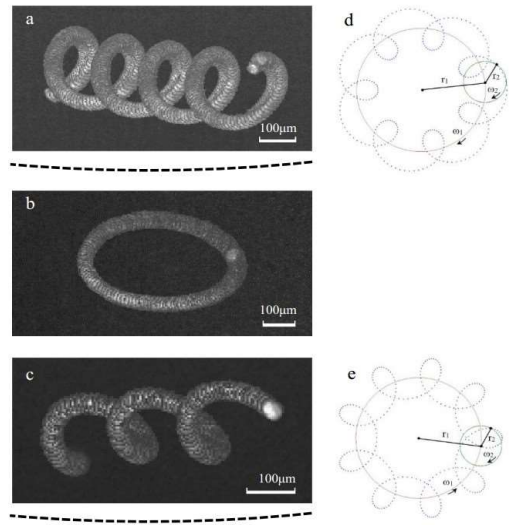

Figure. 2: Trajectories of aggregates. (a): epicycloid motion; (b): elliptical motion; (c): hypocycloid motion. Dashed line indicates schematically the glass ring. (a)-(c) are superpositions of 300 images captured by the high speed camera with the frame rate of 500 fps. (d) and (e) are diagrams of (a) and (c), respectively. The cycloid motions of aggregates follow the equations in polar coordinate if setting the center of the glass ring as the origin of polar coordinate: $r^2 = r_1^2 + r_2^2 + 2r_1 r_2 \cos(\omega_2 - \omega_1)t$, $\theta = \omega_1 t \pm \arccos(\frac{r_1 + r_2\cos(\omega_2 - \omega_1)t}{r})$, as shown by (d)-(e).

For the two kinds of cycloid motions they both satisfy the following relation:

$$|r_1\omega_1| < |r_2\omega_2|, \tag{1}$$

i.e., the linear speed of the secondary circle is less than that of the primary circle. This is a key to the formations of the petals on both epicycloid and hypocycloid trajectories.

The trajectory of aggregate is determined by both the initial conditions (such as the initial position) of the dropped aggregate and the discharge parameters. In general, the aggregate would acquire different initial velocities in radial and/or azimuthal directions while it enters the sheath. This means that for several aggregates dropped simultaneously (or one aggregate dropped repeatedly), they (it) will exhibit different trajectories because their (its) initial conditions are not exactly the same. If we suppose that the direction of the angle velocity of the primary circle points initially upward, i.e., $\omega_2 > 0$, the aggregate which acquires positive initial angular velocity $\omega_1 > 0$ would follows epicycloid trajectory as shown in **figure. 2(a)**. On the contrary, if the aggregate acquires negative initial angular velocity $\omega_1 < 0$, it would follow hypocycloid trajectory as shown in **figure.2(c)**. Therefore, the secondary circle results from the initial angle velocity of the dropped aggregate. In our experiments, the radii of the primary and the secondary circles change within the ranges of $r_2$ =0.08-0.8 mm and $r_1$ =9.5-12 mm, respectively. The corresponding values of angular velocities change within the ranges of $|\omega_2|$=40-90 rad/s and $|\omega_1|$= 0.0-1.4 rad/s, respectively.

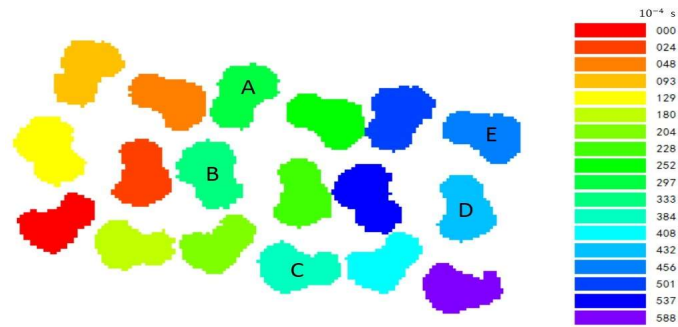

Figure. 3: (Color online) Superposition of images of aggregate which following the epicycloid trajectory at successive moments. The numbers on the colorbar indicate the time in unit of second. The capitals A-E stamped on aggregates correspond to those in Figure. 4.

The primary circle originates from the asymmetry of the aggregate which consists of one large and one small grains. **Figure 3** shows the evolution of an aggregate which following the epicycloid trajectory. It can be seen that in the aggregate the small grain always leads the large one while the aggregate travelling. This means that the aggregate rotates while it travels along the primary circle. The major axis of the aggregate follows approximately the travelling direction of the aggregate. Therefore, the value of the angular velocity of the rotating aggregate is the same as that of the primary circle. As a whole, the aggregate oscillates harmonically near its equilibrium position in the radial direction as shown in **figure. 4(a)**. This also occurs in the cases of hypocycloid and elliptical motions. The leading motion in the aggregate results from the difference of the resonance frequencies of the two grains with different sizes. The resonance frequency of the grain is related with the grain size, $\omega^2 = \frac{Q n_i e}{\epsilon_0 m}$, where, $Q$ is the charge of the grain, $n_i$ is the ion density, $e$ is the elementary charge, $\epsilon_0$ is the vacuum permittivity, $m$ is the mass of the grain. In ref. [18], the relation between the charge $Q$ and the radius $a$ of the grain satisfies $Q \sim a^{1.87}$. Therefore, the resonance frequency $\omega$ is inversely proportional to the radius of the grain. The small grain oscillates faster than the large one, which means that the small grain leads the large one as they travelling.

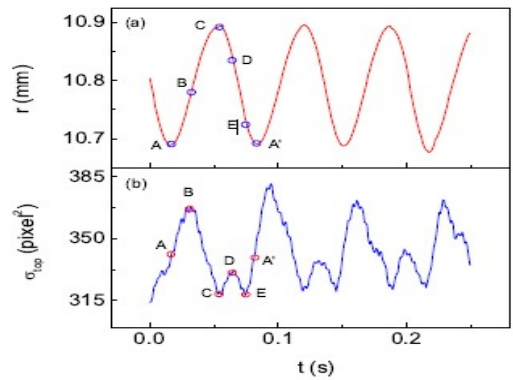

Figure. 4: (Color online) Dependence of the radial position (a) and the cross section $\sigma_{top}$ of top view (b) on time. The capitals A-E correspond to the positions as indicated in Fig. 3.

The force an aggregate suffered during the cycloid motion can be extracted from the recorded trajectory with high spatiotemporal resolution. **Figure. 5** shows the dependence of the radial force on the radial position of an aggregate following hypocycloid trajectory. It exhibits a Lissajou's pattern with the indicated direction. The radial force is of the order of $10^{-12}$ N. It can be seen that the force changes linearly near the equilibrium position $r \approx 10.53$ mm, but is asymmetric while the aggregate oscillates back and forth in radial direction. This should be related with the posture of the aggregate. If we assume that the charge of the aggregate keeps constant due to the small radius $r_2$, then the electrostatic force originated from the action of the sheath field on the negatively-charged aggregate does not change at the same radial position. However, the ion drag force $F_i$ in radial direction from Coulomb collision in the outer sheath region is proportional to the cross section $\sigma_{side}$ (side view) of aggregate which is related with the posture of the aggregate. Therefore, the resultant force $F_r = F_i - F_E$ should be related with the posture of the aggregate. **Figure. 4(b)** shows the evolution of cross section $\sigma_{top}$ of the aggregate from top view, which corresponds to the epicycloid trajectory in **figure. 3** and in **figure. 4(a)**. It can be seen that the average cross section $\sigma_{top}$ from top view as the aggregate travelling outward along A-B-C is larger than that as the aggregate travelling inward along C-D-E-A'. Therefore, the average cross section $\sigma_{side}$ from side view along A-B-C is smaller than that along C-D-E-A'. The slope of the resultant force $F_r$ while the aggregate travelling outward (A-B-C) is smaller than that while the aggregate travelling inward (C-D-E-A'), which gives rise to the Lissajou's pattern as shown in **figure. 5**. In addition, the equilibrium position $r_i$ of the aggregate while it moves outward is smaller than the equilibrium position $r_o$ of the aggregate while it moves inward, which is also related with the posture of the aggregate.

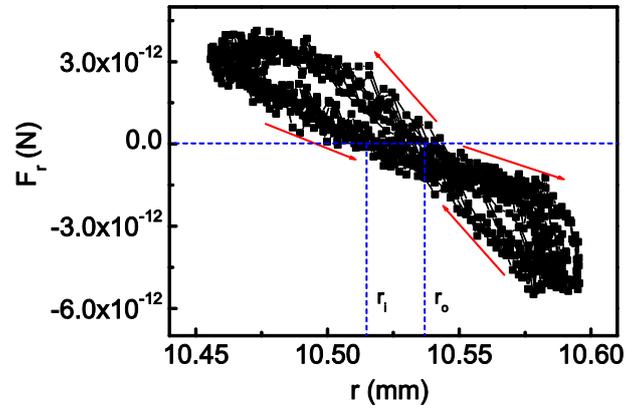

Figure. 5: (Color online) Dependence of the radial force on the radial position of an aggregate following hypocycloid trajectory. It exhibits a Lissajou's pattern with the indicated direction. $r_{i(o)}$ indicates the equilibrium position of the aggregate while it moving outward (inward).

The cycloid motions are the distinctive features of asymmetrical aggregates immersed in a plasma. In order to confirm this, further experiments with glass and resin microspheres are performed at the same discharge conditions for comparison (not shown here). No cycloid motion is observed. Those regular microspheres show behaviors such as the random motion as other groups observed previously. [1, 2]

4. Conclusion

In this work, we have studied experimentally the cycloid motions of aggregates containing one large and one small grains in a dust plasma. The cycloid motions are regarded as combination motions of a primary circle and a secondary circle. Experimental results show that the secondary circle results from the initial angle velocity of dropped aggregate. The primary circle originates from the asymmetry of the aggregate. In the aggregate, the small grain always leads the large one while the aggregate travelling. The cross section of side view changes while the aggregate oscillating back and forth in radial direction, which gives rise to the asymmetry of the resultant force. The dependence of the radial force on the radial position of an aggregate exhibits Lissajou's pattern. Additional comparison experiments with regular microspheres are performed in order to confirm that the cycloid motions are distinctive features of asymmetrical aggregates immersed in a plasma.

**Acknowledgments**

This work is supported by the National Natural Science Foundation of China (Grant No. 11205044, 11405042), the Natural Science Foundation of Hebei Province, China (Grants Nos. A2011201006, A2012201015), the Research Foundation of Education Bureau of Hebei Province, China (Grant No. Y2012009), and the Midwest universities comprehensive strength promotion project.